\documentclass[final, nocrop]{bioinfo}

\usepackage{float}
\usepackage{textcomp}
\usepackage[autolinebreaks,useliterate]{mcode}
\usepackage{linegoal}
\usepackage{array}
\usepackage{multicol}
\usepackage{multirow}
\usepackage{amssymb}
\usepackage[usenames,dvipsnames,svgnames,table]{xcolor}
\usepackage[unicode=true,pdfusetitle, bookmarks=false,breaklinks=true, pdfborder={0 0 0},pdfborderstyle={},backref=false,colorlinks=true]{hyperref}
\hypersetup{colorlinks,citecolor=ForestGreen,filecolor=black,linkcolor=blueTOC,urlcolor=blueTOC}
\usepackage[space]{cite}
\usepackage{listings}
\usepackage{nameref}
\usepackage{etoolbox}

\newsavebox{\mylisting}

\definecolor{mygreen}{RGB}{28,172,0}
\definecolor{mylilas}{RGB}{170,55,241}
\definecolor{mygrey}{RGB}{105,115,126}
\definecolor{myblue}{RGB}{91,146,178}
\definecolor{green_optForce}{RGB}{0, 158, 59}
\definecolor{ForestGreen}{RGB}{34, 139, 34}
\definecolor{yamlGreen}{RGB}{0, 138, 70}
\definecolor{bashRed}{RGB}{240, 92, 63}
\definecolor{bashBlue}{RGB}{37, 95, 172}
\definecolor{blueTOC}{HTML}{1D7DDF}
\definecolor{greenOptSys}{HTML}{75C047}

\lstdefinelanguage{yaml}%
  {morekeywords={language, before_install, script, after_success},%
   sensitive=true,%
   alsoother={$},%
   morecomment=[l]\#,%
   morecomment=[n]{\#=}{=\#},%
   morestring=[s]{"}{"},%
   morestring=[m]{'}{'},%
}[keywords,comments,strings]%

\lstdefinelanguage{Julia}%
  {morekeywords={abstract,break,case,catch,const,continue,do,else,elseif,%
      end,export,false,for,function,immutable,import,importall,if,in,%
      macro,module,otherwise,quote,return,switch,true,try,type,typealias,%
      using,while},%
   sensitive=true,%
   alsoother={$},%
   morecomment=[l]\#,%
   morecomment=[n]{\#=}{=\#},%
   morestring=[s]{"}{"},%
   morestring=[m]{'}{'},%
}[keywords,comments,strings]%

\lstdefinestyle{yamlStyle}{
    language = yaml,
    basicstyle = \footnotesize\ttfamily,
    keywordstyle = \color{yamlGreen},
    stringstyle = \color{magenta},
    commentstyle = \color{mygrey},
    showstringspaces = false,}

\lstdefinestyle{juliaStyle}{
    language = Julia,
    basicstyle = \footnotesize\ttfamily,
    keywordstyle = \bfseries\color{blue},
    stringstyle = \color{magenta},
    commentstyle = \color{ForestGreen},
    showstringspaces = false,}

\lstdefinestyle{bashStyle} {
    language=bash,%
    basicstyle=\footnotesize\ttfamily,%
    breaklines=true,%
    keywordstyle = \bfseries\color{bashBlue},
    stringstyle=\color{bashRed},%
    commentstyle=\color{mygrey},%
    showstringspaces=false,%without this there will be a symbol in the places where there is a space
    emph=[1]{\$},emphstyle=[1]\color{mylilas}, %some words to emphasise
    emph=[2]{cd, git, clone, checkout, pull, fetch, commit, rebase}, emphstyle=[2]\color{myblue},
}

\lstdefinelanguage{XML}
{
  basicstyle=\footnotesize\ttfamily,%
  morestring=[s]{"}{"},
  morecomment=[s]{?}{?},
  morecomment=[s]{!--}{--},
  commentstyle=\color{ForestGreen},
  moredelim=[s][\color{black}]{>}{<},
  moredelim=[s][\color{red}]{\ }{=},
  stringstyle=\color{myblue},
  identifierstyle=\color{bashRed}
}

\begin{document}
\onecolumn

% title
\vspace*{3mm}
\begin{center}
\huge{\texttt{ARTENOLIS}: Automated Reproducibility and Testing\\
Environment for Licensed Software}
\vspace{3mm}
\end{center}

\begin{center}
{\Large{}Laurent Heirendt \& Sylvain Arreckx, Christophe Trefois,
Yohan Yarosz,\\
\vspace*{1mm}
Maharshi Vyas, Venkata P. Satagopam, Reinhard Schneider,\\
\vspace*{2mm}
Ines Thiele, Ronan M.T. Fleming}
\par\end{center}{\Large \par}

\section*{Abstract }

\noindent \textbf{Motivation: }Automatically testing changes to code
is an essential feature of continuous integration. For open-source
code, without licensed dependencies, a variety of continuous integration
services exist. The COnstraint-Based Reconstruction and Analysis (COBRA)
Toolbox is a suite of open-source code for computational modelling
with dependencies on licensed software. A novel automated framework
of continuous integration in a semi-licensed environment is required
for the development of \texttt{the COBRA Toolbox} and related tools
of the COBRA community.

\noindent \textbf{Results:} \texttt{ARTENOLIS} is a general-purpose
infrastructure software application that implements continuous integration
for open-source software with licensed dependencies. It uses a \emph{master-slave
}framework, tests code on multiple operating systems, and multiple
versions of licensed software dependencies. \texttt{ARTENOLIS} ensures
the stability, integrity, and cross-platform compatibility of code
in \texttt{the COBRA Toolbox} and related tools.

\noindent \textbf{Availability and Implementation: }The continuous
integration server, core of the reproducibility and testing infrastructure,
can be freely accessed under \href{http://prince.lcsb.uni.lu/jenkins}{artenolis.lcsb.uni.lu}.
The continuous integration framework code is located in the \mcode{.ci}
directory and at the root of the repository freely available under
\href{http://github.com/opencobra/cobratoolbox}{github.com/opencobra/cobratoolbox}. 

\noindent \textbf{Contact: }\href{mailto:ronan.mt.fleming@gmail.com }{ronan.mt.fleming@gmail.com }

\noindent \textbf{Supplementary information: }Supplementary data are
available at Bioinformatics online.

\noindent \rule[0.5ex]{1\textwidth}{0.5pt}
\begin{multicols}{2}
% --------------------------------------------------------------------------------
\section{Introduction}

Implementation of measures to ensure reproducibility of computational
analyses is a fundamental aspect of scientific credibility {\color{ForestGreen}\citep{Bak16, Mun17}},
in particular in computational biology {\color{ForestGreen}\citep{Bea17}}.
Analysis software developed collaboratively offers the potential for
synergistic effort, but is prone to instability over time due to varying
degrees of specialist domain knowledge of code contributors, laborious
manual integration of individual code contributions, misinterpretation
of the intended operation of code or, especially in academia, the
lack of personnel continuity. Tracking of versions and merging of
code is typically done with version control software, e.g., \texttt{git}
({\color{blueTOC}\href{https://git-scm.com/}{git-scm.com}}).

The process of integrating individual code contributions can be semi-automated
using a \emph{continuous integration} approach {\color{ForestGreen}\citep{Duv07}},
which involves automatically testing proposed software changes before
they are considered for merger with the main code base  {\color{ForestGreen}\citep{Sta14}}. 

Continuous integration enables immediate evaluation of the system-wide
impact of local code changes and accelerates the development of quality
code by enabling the early detection and tracking of errors, defects,
or compatibility issues that can arise when a proposed software change
undermines some previously established functionality in an unanticipated
manner.

Continuous integration systems, such as \texttt{Travis} ({\color{blueTOC}\href{https://travis-ci.org}{travis-ci.org}})
or \texttt{Jenkins} ({\color{blueTOC}\href{https://jenkins.io/}{jenkins.io}}), are
tailored to continuously integrate code and are commonly used for
testing open-source code written in Python {\color{ForestGreen}\citep{Ebr13}}
or Julia {\color{ForestGreen}\citep{Hei17}} on a publicly available infrastructure. 
\texttt{The COBRA Toolbox} is a collaboratively developed software
tool for creation, analysis and mechanistic modelling of genome-scale
biochemical networks {\color{ForestGreen}\citep{HeiArr17}}.
It is distributed as open-source code ({\color{blueTOC}\href{http://github.com/opencobra/cobratoolbox}{github.com/opencobra/cobratoolbox}}), but as it is dependent on MATLAB (The Mathworks, Inc.), using public
infrastructure for continuous integration is hardly possible.  There is a need
for a customisable and fully-controllable environment that offers 
several types of continuous integration jobs, allows for 20
or more concurrent builds, is expandable, permits cross-repository
testing on multiple operating systems (including \emph{macOS}), and
that satisfies short and long-term financial constraints.

\begin{figure*}[t]
\noindent \begin{centering}
\includegraphics[width=0.99\textwidth]{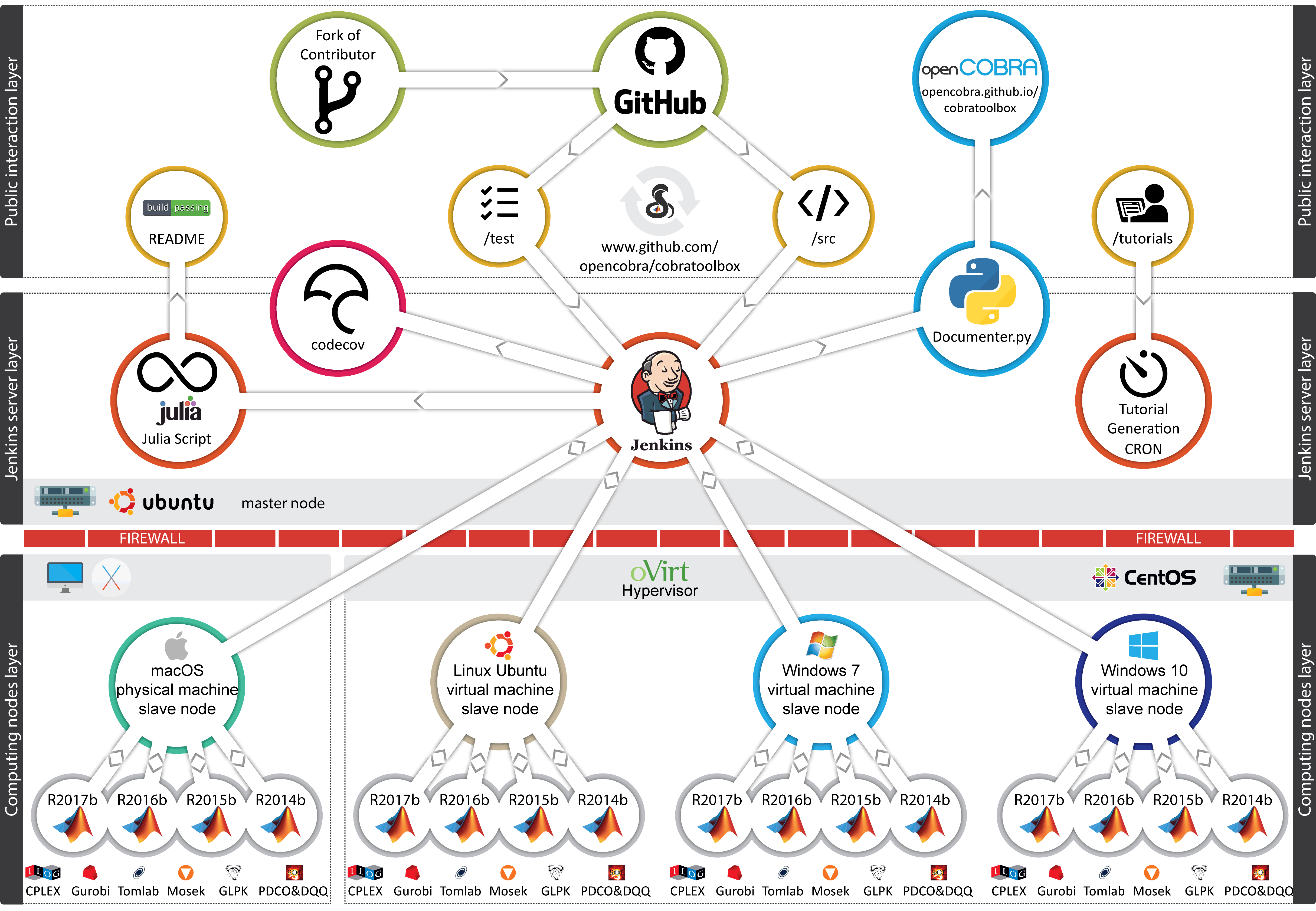}
\vspace*{-0.4cm}
\par\end{centering}
\caption{The core of the infrastructure is \texttt{Jenkins}, the open source
automation server for continuous delivery {\color{ForestGreen}\citep{Fer11}}.
The continuous integration infrastructure consists of a cascade of
3 distinct layers and is a \emph{master-slave} architecture: a \emph{public
interaction layer} (top), the \texttt{\emph{Jenkins}}\emph{ server
layer}, and the \emph{layer with 4 computing nodes} behind a firewall,
each running a different operating system. A change made in a public
repository on \texttt{GitHub} ({\color{blueTOC}\href{https://github.com}{github.com}})
is seen by the \texttt{Jenkins} server (\emph{master}), which in turn
triggers multiple builds on the 4 computing nodes simultaneously (\emph{slaves})
(see the \emph{Supplementary Information} section for details).}
\label{fig:Continuous-integration-setup.}
\vspace{-2mm}
\end{figure*}

This need is apparent when aiming at integrating publicly available code that
is dependent on licensed software, such as MATLAB, and even more so
in regards to technical challenges, such as the activation of licensed
software or the size of proprietary and licensed software container
images.

Here we present a general purpose infrastructure software application
for continuous integration that is compatible with dependencies on
licensed software. We illustrate its utility for development of software
within \texttt{the COBRA Toolbox}, but \texttt{ARTENOLIS} is ready
to be used with other tools using licensed dependencies. This approach ensures that existing
high-quality \texttt{COBRA Toolbox} code is stably maintained to ensure
reproducibility, yet permits the rapid integration of improvements
to existing methods as well as the addition of novel COBRA methods
by an active and geographically dispersed openCOBRA development community
({\color{blueTOC}\href{https://opencobra.github.io/}{opencobra.github.io}}).
% --------------------------------------------------------------------------------
\vspace*{-0.8cm}
\section{Implementation}
\label{sec:Implementation}
\texttt{ARTENOLIS} shown in Figure~\ref{fig:Continuous-integration-setup.}
ensures that the rarely smooth and seamless process of integrating
individual contributions, the so-called \textit{Integration Hell},
is avoided.

The implementation of \texttt{ARTENOLIS} includes the configuration
of \texttt{Jenkins} and its slaves, the continuous integration code
that couples \texttt{ARTENOLIS} to the repository, and customisation
code for repository testing and code quality evaluation. Key to the
present multi-lingual implementation are the unique cross-platform
triggering mechanism, the multiple versions of MATLAB and dependencies,
and the customised and on-the-fly tutorial and documentation deployment.

Prompt feedback on the quality and the stability of the submission
is provided through a comprehensive console output and build badges
before the merger of a code contribution and the evaluation of the
actual stability and the deployment of the documentation. The tutorials
are automatically and regularly tested to ensure error-free execution
with the fast developing code base.

% --------------------------------------------------------------------------------
\vspace{-0.5cm}
\section{Conclusions}
\label{sec:Conclusions}

\texttt{ARTENOLIS} offers a flexible continuous integration solution
for development of any open-source software with licensed dependencies.
Applied to software development in \texttt{the COBRA Toolbox}, it
ensures the stability required for reproducibility of research results
yet accelerates the evolution of software implementations of novel
constraint-based modelling methods.

% --------------------------------------------------------------------------------
\vspace{-0.5cm}
\section*{Acknowledgements and funding}

The authors acknowledge the support from Thomas Pfau for extensively
testing the continuous integration server. The Reproducible Research
Results (R3) team of the Luxembourg Centre for Systems Biomedicine
is acknowledged for support of the project and for promoting reproducible
research.

\texttt{ARTENOLIS} was funded by the National Centre of Excellence
in Research (NCER) on Parkinson's disease, the U.S.
Department of Energy, Offices of Advanced Scientific Computing Research
and the Biological and Environmental Research as part of the Scientific
Discovery Through Advanced Computing program, grant no. DE-SC0010429,
and the Luxembourg National Research Fund (FNR) ATTRACT program (FNR/A12/01)
and OPEN (FNR/O16/11402054) grants.

\noindent \emph{Conflict of Interest:} none declared.

\vspace{-0.5cm}
% --------------------------------------------------------------------------------
\bibliographystyle{bioinformatics}
\fontsize{7}{9}\selectfont 

\end{multicols}

\onecolumn

% title
\vspace*{3mm}
\begin{center}
\huge{Detailed continuous integration setup}
\vspace{2mm}

\LARGE{Supplementary information}
\vspace{3mm}

\vspace{3mm}
\end{center}

% define the labels of the figures and tables
\setcounter{section}{0}
\setcounter{page}{1}
\let\oldthetable\thetable 
\renewcommand{\thetable}{S\oldthetable}
\let\oldlstlisting\thelstlisting
\renewcommand{\thelstlisting}{S\oldlstlisting}
\let\oldfigure\thefigure
\renewcommand{\thefigure}{S\oldfigure}
\let\oldsection\thesection
\renewcommand{\thesection}{S\oldsection}
% start of content
\normalsize

\begin{multicols}{2}
% --------------------------------------------------------------------------------
\section{Background and overview}

Software engineers in industry routinely apply continuous integration
approaches to ensure that code developed in a collaborative way results
in stable and cross-platform compatible software products \citep{Sil17}
and to shorten turnaround times. Continuous integration is used more
and more often in computational biology, especially for reproducing
data-driven research results \citep{Gar13,Nar16},
but rarely when software has a dependency on commercial or licensed
software. 

Continuous integration practices are adopted at a different pace for
industrial and open-source software projects as both face different
challenges in software development \citep{Hol03}.
Typically, in a research environment, code might not be used beyond
the scientist who authored the code. Researchers focus on obtaining
results swiftly rather than writing prototype code that is compatible
with several licensed software versions or operating systems. A common
practice is to store multiple versions of code on various media with
regular backups. This setup eventually works well for a small code
base used by a single user, but is prohibitive in a collaborative
environment, and more so when intending to guarantee reproducibility
of results and aiming to provide a stable code base. 

According to \citet{Duv07} and \citet{Fow06} 
who analysed similar situations, a continuous integration system should
centre itself in a development process aiming for stability, high
quality, reproducible results, bug-free code execution, and a pleasant
end-user experience. Metrics for code quality, such as code grade
or code coverage, help developing functional, high-quality code, and
ultimately, code that ensures reproducibility of results (see Section
\ref{subsec:Code-quality}).

For growing projects, scaling up the continuous integration system
could be envisaged \citep{Mey14}. Nonetheless, while
the code repositories that are continuously integrated must be maintained,
the same holds for a running environment for reproducibility and testing
\citep{Rog04}.

Integrated software development and analysis environments, such as
MATLAB, make it straightforward to start coding and modify existing
code, even for novice users. Within the COBRA community, end users
run their code on a variety of operating systems, such as \emph{Windows},
\emph{macOS,} or \emph{Linux}. \texttt{ARTENOLIS} benefits all end
users considerably by ironing out incompatibilities and defects, especially
\emph{Windows} users. The underlying server architecture is explained
in detail in Section \ref{subsec:Server-architecture} and the configuration
of \texttt{ARTENOLIS} is described in Section \ref{sec:Configuration-of-Continuous}. 

\vspace*{-2mm}
% --------------------------------------------------------------------------------
\section{Cascade of the Continuous Integration system}
\label{sec:Continuous-integration-system}

The cascade shown in Figure~1
is typical of a continuous integration system of a reproducibility
and testing infrastructure, but includes specificities tailored to
\texttt{the COBRA Toolbox}.

% --------------------------------------------------------------------------------
\subsubsection*{Public interaction layer}

The main code base, as well as all of its \emph{forks}, are located
on GitHub, e.g., the repository of \texttt{the COBRA Toolbox} is located
at \href{http://github.com/opencobra/cobratoolbox}{github.com/opencobra/cobratoolbox}.
A \emph{fork} is an individual and modifiable copy of the main code
base of the contributor with \emph{<userName>}, and is located under
\href{http://github.com/<userName>/cobratoolbox}{github.com/<userName>/cobratoolbox}.
Contributors may contribute to the main code base by opening a \emph{pull
request }(PR), which is reviewed and tested. In particular, developers
of \texttt{the COBRA Toolbox} repository may easily submit pull requests
using the\emph{ }\texttt{MATLAB.devTools} (\href{https://github.com/opencobra/MATLAB.devTools}{github.com/opencobra/MATLAB.devTools}).

The public interaction layer also includes the badges of the build
status (see Section \ref{subsec:Build-status}) and the publicly accessible
website \href{http://opencobra.github.io/cobratoolbox}{opencobra.github.io/cobratoolbox},
which hosts the documentation and tutorials. The details on the generation
and deployment of the documentation are provided in Section \ref{sec:Documentation-and-tutorials}.
The repository of \texttt{the COBRA Toolbox} is structured according
to common practices of continuous integration (e.g., \href{https://github.com/JuliaLang/julia}{github.com/JuliaLang/julia})
and is explained in detail in Section \ref{subsec:Repository-structure-and}.

% --------------------------------------------------------------------------------
\subsubsection*{Jenkins server layer}

The main element of the\texttt{ Jenkins} server layer is the \texttt{Jenkins}
software installed on a virtual machine, which acts as a \emph{master}
node. The\texttt{ Jenkins} server is accessible from the Internet
and is constantly listening for activity on the main GitHub repository
via \emph{web-hooks}. Once a PR is submitted by a contributor, builds
are triggered on the slave machines through secure Java Web Start
communication. 

The\texttt{ Jenkins} server layer partially includes the \texttt{codecov}
and \texttt{Documenter.py} elements. The coverage report is prepared
on the\texttt{ Jenkins} server, and is being made publicly available
through \texttt{codecov} under \href{http://codecov.io/gh/opencobra/cobratoolbox}{codecov.io/gh/opencobra/cobratoolbox}.
Similarly, the documentation is generated on the\texttt{ Jenkins}
server before being deployed to the \href{http://opencobra.github.io/cobratoolbox}{opencobra.github.io/cobratoolbox}
website.

% --------------------------------------------------------------------------------
\subsubsection*{Computing nodes layer}

The computing nodes layer contains 4 computing nodes, each running
a different operating system. Most importantly, the continuous integration
system set up for \texttt{the COBRA Toolbox} runs on the most popular
operating systems in the COBRA community: \emph{Linux}, \emph{macOS},\emph{
Windows 7}, and \emph{Windows 10}. A virtualisation environment bears
key advantages for a continuous integration system and is described
in Section \ref{subsec:Server-architecture}. 

When triggered, each of the 4 computing nodes launches the last 4
stable versions of MATLAB (R2014b, R2015b, R2016b, and R2017b). Each
MATLAB version then runs the dedicated testing suite (see Section
\ref{subsec:Continuous-integration-test}), which makes use of multiple
solvers, such as CPLEX \citep{Ibm17}, \citet{Gur17},
Tomlab \citep{Hol04}, Mosek \citep{And00},
GLPK (GNU Linear Programming Kit), PDCO \citep{Che98}
and DQQ \citep{Ma17}.

\vspace*{-2mm}
% --------------------------------------------------------------------------------
\section{Architecture }
\label{subsec:Server-architecture}

% --------------------------------------------------------------------------------
\subsection{Computing nodes and resources}
\label{subsec:Computing-nodes-and}

\begin{table*}[t]
\noindent \begin{centering}
\small{%
\begin{tabular*}{1\textwidth}{@{\extracolsep{\fill}}>{\centering}m{1.5cm}>{\centering}m{2.8cm}ccc>{\centering}m{1.5cm}>{\centering}m{2.1cm}>{\centering}m{2.2cm}}
\textbf{Physical machine } & \textbf{Name} & \textbf{Type} & \textbf{Mode} & \textbf{Memory} & \textbf{Number of cores} & \textbf{Operating System} & \textbf{Storage}\tabularnewline
\hline 
1 & \emph{Jenkins server} & Virtual & master & 8 GB ECC & 4 & \emph{Ubuntu 16.04} & 34GB (OS) + 120GB (data)\tabularnewline
\hline 
\multirow{3}{1.5cm}{\centering2} & \emph{Linux node} & \multirow{3}{*}{Virtual} & \multirow{3}{*}{slave} & \multirow{3}{*}{18 GB ECC} & \multirow{3}{1.5cm}{\centering20} & \emph{Ubuntu 16.04} & 46GB (OS) + 200GB (data)\tabularnewline
\cline{2-2} \cline{7-8} 
 & \emph{Windows 7 node} &  &  &  &  & \emph{Windows 7} & 60GB (OS) + 200 GB(data)\tabularnewline
\cline{2-2} \cline{7-8} 
 & \emph{Windows 10 node} &  &  &  &  & \emph{Windows 10} & 50 GB (OS) + 200 GB (data)\tabularnewline
\hline 
3 & \emph{Mac node} & Physical & slave & 24 GB ECC & 12 & \emph{macOS 10.12} & 250 GB\tabularnewline
\hline
\end{tabular*}}
\par\end{centering}
\vspace{1mm}
\caption{Technical specifications of \emph{master }and\emph{ slave} nodes (computing
nodes layer) within \texttt{ARTENOLIS}.}
\label{tab:Specifications-of-machines}
\end{table*}

The advantage of virtual machines over physical machines is that various
computing environments can co-exist on a single physical machine.
A setup with fewer physical machines but with multiple virtual environments
is more economical and occupies a smaller space in the server racks.
Thanks to a virtualisation layer and a hypervisor monitoring the health
(see Section \ref{subsec:Virtualization-layer}), new virtual machines
can be created or deleted on demand and based on the available capacity
of the physical server. The \texttt{Jenkins} virtual machine running
the \emph{Linux Ubuntu} operating system is a shared resource, is
referred to as the \emph{master} node\emph{,} handles HTTPs requests,
and orchestrates the slave nodes. 

The \emph{Linux} (Debian based), \emph{Windows 7,} and \emph{Windows
10} operating systems are virtual machines and are running on the
same physical computing node, whereas the \emph{macOS} operating system
is running on a dedicated physical machine. The specifications of
the computing nodes are provided in Table \ref{tab:Specifications-of-machines}.
A limited amount of storage is usually provided to each virtual machine
and only satisfies the need to run and store small amount of data.
For larger data, such as build data, an NFS (Network File System)
or SMB (Server Message Block) mount point is provided on a central
storage system. 

% --------------------------------------------------------------------------------
\subsection{Virtual machine management and access control}
\label{subsec:Virtual-machine-management}
Security and appropriate access control to each virtual machine and
associated services is provided through \emph{FreeIPA} (\href{http://freeipa.org}{freeipa.org}).
\emph{FreeIPA} provides a centralised resource to control authentication,
authorisation and account information by storing all information about
a user, virtual machine and other sets of objects required to manage
the security of the virtual machine.

In order to reduce maintenance and initialisation time of new virtual
machines, the virtual machines are deployed using \emph{Foreman} (\href{http://theforeman.org}{theforeman.org}),
a virtualisation environment agnostic web tool typically used to manage
the complete lifecycle of virtual machines. 

All physical and virtual servers (except the \emph{macOS} node) are
configured via \emph{Puppet}, which is a configuration management
tool that facilitates standardised configurations across a pool of
servers such as firewall definitions, administration SSH (Secure Shell)
keys, default packages, and other settings. In the current setup,
the compliance of all machine configurations is constantly monitored,
while periodic health reports are provided to \emph{Foreman.} Besides
the configuration monitoring and deployment, the performance of the
virtual machines is monitored using \emph{netdata} (\href{https://github.com/firehol/netdata}{github.com/firehol/netdata}),
which is particularly useful when evaluating the performance of the
continuous integration test suite (see Section \ref{subsec:Job-definitions}).

% --------------------------------------------------------------------------------
\subsection{Virtualisation layer}
\label{subsec:Virtualization-layer}
Hypervisor software runs on a server handling virtual machines (VMs).
In the present case, \emph{oVirt} (\href{http://ovirt.org}{ovirt.org})
is installed on each of the physical servers, and is a key element
in VM management. \emph{Kernel-based Virtual Machine} (KVM), a virtualisation
infrastructure,\emph{ }is used as the hypervisor layer. \emph{oVirt}
provides a graphical user interface (GUI) to manage all physical and
logical resources needed for the virtual infrastructure (e.g storage,
network, data centres). In the current continuous integration system,
the 2 physical servers (not the \emph{macOS} node, see Table \ref{tab:Specifications-of-machines})
are running \emph{CentOS 7.3} and are virtualised using \emph{oVirt
4.1.0}. 

\vspace{-3mm}
% --------------------------------------------------------------------------------
\section{Configuration ARTENOLIS}
\label{sec:Configuration-of-Continuous}
% --------------------------------------------------------------------------------
\subsection{Repository structure and branches}
\label{subsec:Repository-structure-and}

In Table \ref{tab:Directory-structure-of}, the main directories of
\texttt{the COBRA Toolbox} repository are listed together with their
respective purpose. The test suite is located in \emph{/test}, while
the source files of the code base are located in the \emph{/src} directory.
These two directories, together with \emph{/tutorials}, are key components.
The \emph{/.ci }directory contains scripts required for the continuous
integration integration

The scripts \emph{travis.yml }and \emph{codecov.yml} located at the
root of the repository are required to trigger the jobs (see Section
\ref{subsec:Continuous-integration-(CI)}) and to report code coverage
(see Section \ref{subsec:Code-coverage}), respectively.
\vspace{-2mm}
\begin{table}[H]
\noindent \begin{centering}
\small{%
\begin{tabular*}{1\columnwidth}{@{\extracolsep{\fill}}l>{\raggedright}m{5cm}}
\textbf{Directory/file name} & \textbf{Purpose}\tabularnewline
\hline 
\emph{/.ci} & Directory with continuous integration bash-scripts \tabularnewline
\hline 
\emph{/src} & Directory with code source files\tabularnewline
\hline 
\emph{/test} & Directory with test files and \emph{testAll.m} \tabularnewline
\hline 
\emph{/tutorials} & Directory with tutorial \emph{.mlx} files \tabularnewline
\hline 
\emph{travis.yml} & YAML trigger script\tabularnewline
\hline 
\emph{codecov.yml} & YAML script for code coverage report \tabularnewline
\hline 
\end{tabular*}}
\par\end{centering}
\vspace{1mm}
\caption{Structure and key directories and files of \texttt{the COBRA Toolbox} repository.}
\label{tab:Directory-structure-of}
\end{table}
\vspace{-5mm}
As explained in Section \ref{subsec:Job-definitions}, \texttt{the
COBRA Toolbox} follows the common development model of a stable \emph{master}
branch and a \emph{develop} branch for development. This development
model is particularly well suited for a reproducibility and testing
infrastructure, such as\texttt{ ARTENOLIS}. It is against the \emph{develop}
branch that new pull requests are raised by developers, while it is
the \emph{master} branch that contains the stable version of\texttt{
the COBRA Toolbox}. A regular merge strategy from the \emph{develop}
to the \emph{master }branch ensures that the latest features are adopted
in the stable version. As the development of new features is made
on separate branches that are being merged to the \emph{develop} branch
only through Pull Requests (PRs), and as each PR is being tested by
the continuous integration system, the risk of the \emph{develop}
branch failing to build is very low. This stability of the \emph{master}
branch is particularly important in a fast-moving research environment
that relies on the reproducibility of data-driven results.

\vspace{-2mm}
% --------------------------------------------------------------------------------
\subsection{Local and GitHub user accounts}
\label{subsec:Local-and-Github}

\begin{table*}[t]
\noindent \begin{centering}
\small{%
\begin{tabular*}{1\textwidth}{@{\extracolsep{\fill}}l>{\raggedright}p{3.5cm}cccc}
\textbf{Job name} & \textbf{Description} & \textbf{R2017b} & \textbf{R2016b} & \textbf{R2015b} & \textbf{R2014b}\tabularnewline
\hline 
COBRAToolbox-branches-auto-\textbf{linux} & \multirow{4}{3.5cm}{Build the \emph{develop} and \emph{master} branches (automatic trigger)} & $\star$ & $\star$ & $\star$ & $\star$\tabularnewline
\cline{1-1} \cline{3-6} 
COBRAToolbox-branches-auto-\textbf{macOS} &  & $\star$ & $\star$ & $\star$ & $\star$\tabularnewline
\cline{1-1} \cline{3-6} 
COBRAToolbox-branches-auto-\textbf{windows7} &  & $\star$ & $\star$ & $\star$ & $\star$\tabularnewline
\cline{1-1} \cline{3-6} 
COBRAToolbox-branches-auto-\textbf{windows10} &  & $\star$ & $\star$ & $\star$ & $\star$\tabularnewline
\hline 
\hline 
COBRAToolbox-pr-auto-\textbf{linux} & \multirow{4}{3.5cm}{Build any newly submitted Pull Request } & $\star$ & $\star$ & $\star$ & $\star$\tabularnewline
\cline{1-1} \cline{3-6} 
COBRAToolbox-pr-auto-\textbf{macOS} &  &  & $\star$ &  & \tabularnewline
\cline{1-1} \cline{3-6} 
COBRAToolbox-pr-auto-\textbf{windows7} &  &  & $\star$ &  & \tabularnewline
\cline{1-1} \cline{3-6} 
COBRAToolbox-pr-auto-\textbf{windows10} &  &  & $\star$ &  & \tabularnewline
\hline 
\hline 
COBRAToolbox-branches-manual-linux & Build the \emph{develop} and \emph{master} branches by SHA1 (Secure
Hash Algorithm 1)\\
\textit{(manual trigger)} & $\star$ & $\star$ & $\star$ & $\star$\tabularnewline
\hline 
COBRAToolbox-pr-manual-linux & Build a Pull Request by SHA1 \textit{}\\
\textit{(manual trigger)} & $\star$ & $\star$ & $\star$ & $\star$\tabularnewline
\hline 
\end{tabular*}}
\par\end{centering}
\vspace{1mm}
\caption{Job definitions for the continuous integration setup of \texttt{the
COBRA Toolbox} repository.\label{tab:Job-definitions-for}}
\end{table*}

A local non-administrative user account on the slave nodes is set
up in order to run MATLAB as a specific user during a continuous integration
build, have proper read/write permissions, and allow for simplified
repository maintenance and/or debugging. This user is not bound to
a physical administrator, and can be associated to licenses or perform
repetitive tasks, such as setting the build badges (see Section \ref{subsec:Continuous-integration-feedback}).
This local user account is also not intended to be used to perform
administrative tasks on the computing node. Within \texttt{ARTENOLIS},
this local user account is named \emph{jenkins}, and is independent
of server-wide access control as explained in Section \ref{subsec:Virtual-machine-management}.

In order to perform certain repository related tasks, such as deploying
the documentation (see Section \ref{sec:Documentation-and-tutorials}),
a dedicated GitHub account that is not related to a physical person
has been created. This GitHub account is a bot-type account, named
\emph{cobrabot}, and is only used for administrative tasks or committing/pushing
automatically to the GitHub server.

% --------------------------------------------------------------------------------
\subsection{\emph{master} and \emph{slave} nodes}
\label{subsec:Configuration-of-master}

The configuration of the continuous integration system exploits the
\emph{master/slave} functionality integrated in \texttt{Jenkins}.
A large number of jobs running on various operating systems (see Section
\ref{subsec:Job-definitions}) may be triggered from a single \texttt{Jenkins}
server. The workload of multiple jobs can be delegated by the \emph{master}
node to multiple \emph{slave} nodes, which allows for a single \texttt{Jenkins}
installation. The \emph{master} node serves the web interface of \texttt{Jenkins}
(\href{http://artenolis.lcsb.uni.lu/}{artenolis.lcsb.uni.lu})
and acts as a portal to the entire farm of slave nodes. The load of
the jobs is distributed on the \emph{master }node, while the \emph{slave}
nodes are triggered accordingly.

The slaves are connected to the master node through Java Web Start
(Java Network Launch Protocol - JNLP). An agent is running on the
slave that listens for a triggering signal from \texttt{Jenkins} on
the master node. Once triggered, the \emph{slave} node executes the
job and reports back to the \emph{master }node a build status as explained
in Section \ref{subsec:Build-status}. Importantly, on all computing
nodes, the service running the agent must be configured to launch
upon startup of the server, as \texttt{Jenkins} on the \emph{master}
node is not able to wake the agent on the slaves. On \emph{macOS},
it is important to run \emph{Caffeine} (\href{http://lightheadsw.com/caffeine}{lightheadsw.com/caffeine}),
a tiny program that prevents the \textit{macOS} system from automatically
activating the sleep function.

% --------------------------------------------------------------------------------
\subsection{Job definitions}
\label{subsec:Job-definitions}
A job is a configuration of a build pipeline on the \texttt{Jenkins}
server, has a specific purpose, runs on a different slave/operating
system, or builds a different branch of the repository. An example
of such a configuration file is given under \href{http://artenolis.lcsb.uni.lu/userContent/configExample.yml}{artenolis.lcsb.uni.lu/userContent/configExample.yml}.
The job definitions in \texttt{Jenkins} on the master node are configured
accordingly as indicated in Table \ref{tab:Job-definitions-for}. 

In essence, one job is defined per operating system, and in this particular
continuous integration setup, one job per slave. This setup has been
chosen in order to provide the highest robustness of \texttt{ARTENOLIS}.
A \texttt{Jenkins} job may also be parameterised. In the case of \texttt{the
COBRA Toolbox} repository, a matrix of sub-jobs is generated for different
MATLAB versions using the \mcode{MATLAB_VER} parameter.

In order to streamline the continuous integration setup, there are
two distinct job types: jobs that trigger builds of the \emph{develop}
and \emph{master} branches (marked with \mcode{-branches}) and jobs
marked with \mcode{-pr} that build any newly submitted pull request
(PR). 

Each job type is either triggered automatically by \texttt{Jenkins}
or manually by an administrator using the SHA1 of the commit. As the
\emph{develop} and \emph{master }branches must be tested for all supported
MATLAB versions, the \mcode{-branches} jobs run for each supported
MATLAB version on each slave. However, whenever a Pull Request is
submitted, only the job running on the \emph{Linux} node triggers
the test suite on all supported MATLAB versions. On all other slaves
(\emph{macOS}, \emph{Windows} \emph{7}, and \emph{Windows 10}), only
the most stable and most used version of MATLAB is tested. This setup
of the reproducibility and testing infrastructure is tailored to reduce
the required computational resources to a minimum.

In order to ensure flexibility of the continuous integration system
and in case of emergency, the same job types may be manually triggered
on the \textit{Linux} platform. Whenever the \emph{develop} or \emph{master}
branches failed to build, or pull requests failed to be triggered
properly, the specific SHA1 of the commit failing the branch may be
built separately. In addition, the job may also be manually relaunched.
The manual trigger is only available on the \emph{Linux} node in order
to permit swift debugging, rapid intervention, and lowest resource
consumption. Another good practice is to configure a job for testing
purposes (e.g., a job that only builds a specific branch) before making
any production changes on critical jobs (i.e., the \mcode{-branches}
jobs).

Although \emph{Mathworks Inc.} releases two versions of MATLAB per
year, only the last released versions (version \emph{b}) are officially
supported by \texttt{the COBRA Toolbox} and run with various solver
packages on each supported operating system. Certain solvers are on
a different release schedule, and the compatibility of certain solvers
is generally only ensured a long time after the release of a new MATLAB
version. These combinations are prone for incompatibilities. The compatibility
of \texttt{the COBRA Toolbox} has been tested and evaluated separately
such that incompatibility issues between solver versions, operating
systems, and MATLAB versions do not crash on the continuous integration
server or on the user system. A solver compatibility matrix has been
established for all supported operating systems and actively supported
solver interfaces with their respective versions. This compatibility
matrix is used by MATLAB during runtime to determine whether a certain
user setup is compatible or not (\href{http://opencobra.github.io/cobratoolbox/docs/compatibility.html}{opencobra.github.io/cobratoolbox/docs/compatibility.html}),
and ensures that certain tests on the continuous integration setup
are skipped, which would otherwise lead to a build failure.

Each job can either succeed or fail. The build stability of each job
and build trend is monitored and can be retrieved from \href{http://artenolis.lcsb.uni.lu/job/<jobName>/buildTimeTrend}{artenolis.lcsb.uni.lu/job/<jobName>/buildTime-Trend},
where \emph{<jobName>} is the name of a job as listed in Table \ref{tab:Job-definitions-for}.
Currently, a job on the continuous integration system takes in average
around 30-40 minutes to finish (all MATLAB versions). Although all
MATLAB versions are launched in parallel, and despite some tests of
the test suite requesting a parallel pool of workers, the memory consumption
is moderate. Each job requests about 12GB of memory, while all CPUs
of the virtual machines are used of up to 60\%. For each job, input/output
(IO) on the slaves is high at the beginning during the repository
cloning phase, but is negligible during the test run itself. If more
CPU power or RAM was needed in the future, the technical characteristics
of the virtual machines may effortlessly be changed, which is another
advantage of the \emph{master-slave} setup and of the virtual machines
as explained in Sections \ref{subsec:Configuration-of-master} and
\ref{subsec:Computing-nodes-and}. As shown in Table \ref{tab:Specifications-of-machines},
the hard-drive of each of the slaves has limited capacity. All builds
are stored on a central storage server, and old builds are discarded.
The workspace itself is cleaned after each job in order to reduce
the storage needs.

The performance of the slaves and of the hypervisor are monitored
internally using \emph{netdata} (\href{http://github.com/firehol/netdata}{github.com/firehol/netdata}).
Over time, the \texttt{Jenkins} memory usage on the \emph{master }node
increases gradually. In order to avoid the \emph{master }node to swap
memory, \texttt{Jenkins} is restarted every night when being idle.
A regular restart also ensures that the configuration files of \texttt{Jenkins}
are reloaded and that the latest configuration files of \texttt{Jenkins}
and the jobs are used.

% --------------------------------------------------------------------------------
\subsection{Continuous integration scripts}
\label{subsec:Continuous-integration-(CI)}

When triggered, each job starts by cloning the repository, and checks
out the latest commit or the commit that shall be built. On the \texttt{Jenkins}
node, the Java web call triggers the interpretation of a YAML (YAML
Ain't Markup Language) script, namely the \emph{travis.yml
}script shown in Listing \ref{lst:travisScript}. Together with build-specific
environment variables, an executable shell script, also known as the
Hudson shell file, is generated. This Hudson shell file is then sent
to and executed on each slave in a shell-like environment. The continuous
integration process is consequently started through the \emph{travis.yml}
script placed at the root of the repository directory. 

\begin{lstlisting}[caption={.travis.yml script interpreted by Jenkins and used to trigger the \mcode{runtests.sh} script.}\vspace{1mm},label={lst:travisScript}, style=yamlStyle, frame=single, basicstyle=\small\ttfamily]
language: bash

before_install:
	# fresh clone of the repository
	- if [[ -a .git/shallow ]]; then 
		git fetch --unshallow; 
	  fi

script:   
    # launch the tests
	- bash .ci/runtests.sh      
\end{lstlisting}
\vspace{3mm}

As a shell script offers cross-platform flexibility, a proper shell
script is called from the YAML script, namely \emph{runtests.sh} located
in the specific \emph{/.ci }folder. The shell script \emph{runtests.sh}
is shown in Listing \ref{lst:runtestsScript} and runs on all supported
platforms. Each platform is identified by the environment variable
\mcode{ARCH}, which is set by the job definitions (see Section \ref{subsec:Job-definitions}).
The simplest launch command is set for UNIX operating systems (\emph{Linux} and \emph{macOS}). 

As explained in Section \ref{subsec:Configuration-of-master}, the
usage of the tiny \emph{caffeine }program is obvious when running
the script on \emph{macOS}. On UNIX, any output of MATLAB while running
the \emph{testAll.m} script is routed directly to the shell of the
slave, and ultimately, to the console publicly accessible on \texttt{Jenkins}
under \href{http://artenolis.lcsb.uni.lu/job/<jobName>/<buildNumber>/MATLAB_VER=<MATLABversion>,label=<platform>/console}{artenolis.lcsb.uni.lu/job/<jobName>/<buildNumber>/MATLAB\_{}VER=\\<version>,label=<platform>/console},
where \emph{<jobName>} is the name of the job as defined in Table
\ref{tab:Job-definitions-for}, \emph{<buildNumber>} is the number
of the build, \emph{<platform>} is the label of the operating system,
and \emph{<version>} is the MATLAB version.

\begin{lstlisting}[caption={\mcode{runtests.sh} script used to launch the MATLAB sessions}\vspace{1mm}, label={lst:runtestsScript}, style=bashStyle, frame=single, float=*, basicstyle=\small\ttfamily]
#!/bin/sh 
if [ "$ARCH" == "Linux" ]; then  
	$INSTALLDIR/MATLAB/$MATLAB_VER/bin/./matlab -nodesktop -nosplash < test/testAll.m

elif [ "$ARCH" == "macOS" ]; then     
	caffeinate -u &    
	$INSTALLDIR/MATLAB_$MATLAB_VER.app/bin/matlab -nodesktop -nosplash < test/testAll.m

elif [ "$ARCH" == "windows" ]; then     
	# change to the build directory    
	echo " -- changing to the build directory --"   
	cd "D:\\jenkins\\workspace\\COBRAToolbox-windows\\MATLAB_VER\\$MATLAB_VER\\label\\$ARCH"

    echo " -- launching MATLAB --"  
	unset Path     
	nohup "D:\\MATLAB\\$MATLAB_VER\\\bin\\matlab.exe" -nojvm -nodesktop -nosplash -useStartupFolderPref -logfile output.log -wait -r "restoredefaultpath; cd test; testAll;" & PID=$!

    # follow the log file     
	tail -n0 -F --pid=$! output.log 2>/dev/null

    # wait until the background process is done    
	wait $PID 
fi

CODE=$? 
exit $CODE
\end{lstlisting}

A key challenge with setting up \texttt{ARTENOLIS} is to trigger a
build on a \emph{Windows} platform and launch MATLAB while providing
live feedback, similar to UNIX platforms. As no native \emph{bash}
console exists on DOS-based systems, the output of MATLAB is not directly
routed to the console. Instead, \texttt{git Bash} (\href{http://git-scm.com/download/win}{git-scm.com/download/win})
is used, and the Hudson script launching MATLAB is executed in \emph{sh.exe}.
As the output from a shell-like environment is not routed back to
the \emph{master} node directly, a computational trick must be used
in order to however display a live console output on the \texttt{Jenkins}
web interface. This trick ensures the homogeneity of \texttt{ARTENOLIS}
despite the large differences between DOS and UNIX operating systems
supported by \texttt{ARTENOLIS}.

The output of MATLAB is routed to a file (\emph{output.log}), while
the process is run as a hidden process marked with \mcode{&}. The
process ID is saved as \mcode{PID}. While the hidden process is running,
the log file is constantly read in (or \emph{followed}) by the system
command \mcode{tail}, whose output is redirected to \texttt{Jenkins}.
The shell script \emph{runtests.sh} then only exits once the MATLAB
process with \mcode{PID} has been executed without an error. 

Any eventual error code thrown during this process is caught in the
variable \mcode{$CODE}. This is the exit code of the script \emph{runtests.sh}
that is returned to \texttt{Jenkins} running on the \textit{master}
node. More details on how this feedback code is interpreted are given
in Section \ref{subsec:Continuous-integration-feedback}.

% --------------------------------------------------------------------------------
\subsection{GitHub interaction}
\label{subsec:GitHub-interaction}

In order for a build to be triggered from GitHub on the \texttt{Jenkins}
server, a so-called \emph{web-hook} has been installed on GitHub. This
hook listens to events that occur on the GitHub repository of \texttt{the
COBRA Toolbox} and include the opening of a pull request, a change
of a commit, or another status modification. On \texttt{Jenkins},
a cookie is stored, which allows \texttt{Jenkins} to listen to that
particular hook and that cannot be triggered by any other hook.

A valuable feature of the continuous integration system is that the
continuous integration software \texttt{Jenkins} integrates seamlessly
with the version control server GitHub. Next to each commit
that has triggered a build on the continuous integration system, a
status is displayed (see Section \ref{subsec:Build-status}). This visual 
system is standard for continuously integrated
repositories on GitHub, and allows developers to swiftly check and
track the status of a build from GitHub directly.

% --------------------------------------------------------------------------------
\section{Evaluation of code stability}
\label{sec:MATLAB-testsuite-and}

% --------------------------------------------------------------------------------
\subsection{Continuous integration test suite}
\label{subsec:Continuous-integration-test}

The test-suite for relies on testing framework functions and is designed
to test the functionality and performance of the MATLAB code of \texttt{the
COBRA Toolbox}. Once MATLAB is running on the continuous integration
server, the test-suite is launched from the \mcode{testAll} script.
The test-suite makes use of the dedicated unit testing functions implemented
in MATLAB. The \mcode{testAll} script is structured such that \texttt{MOcov}
and \texttt{JsonLab} are added to the path first, global variables
are defined, and the code quality grade is determined (see Section
\ref{subsec:Code-quality-grade}) before running any unit-test functions.
The \mcode{runtests} command runs in serial (or in parallel) all
test files in the \emph{/test} folder. Once the \mcode{runtests}
command completed, the number of tests that failed and/or are incomplete
is evaluated, and the code coverage percentage is computed (see Section
\ref{subsec:Code-coverage} for details).

A test, part of the continuous integration test suite, is tailored
to run a function in the \emph{/src} directory such that a result
is output, a warning, or an error are thrown. A test must evaluate
the result returned by the function against a pre-computed reference
result. This evaluation within the test is performed using the \mcode{assert}
function. All tests for \texttt{the COBRA Toolbox} follow a template,
and a guide helps developers get started (\href{http://opencobra.github.io/cobratoolbox/docs/contributing.html}{opencobra.github.io/cobratoolbox/docs/contributing.html}).

The \mcode{runtests} command prints a table with the test name, the
running time of each test, and whether the test failed or succeeded.
As the exit status code returned by \mcode{runtests} determines the
exit code of the MATLAB process (see Section \ref{subsec:Continuous-integration-(CI)}),
the continuous integration test suite is launched within a \mcode{try...catch}
statement. The exit code is explicitly set to \mcode{1} if the
number of failed or incomplete tests is not $0$. The \mcode{exit(exit_code)}
command is called on the continuous integration server before the
end of the \mcode{try...catch} statement. Any exception thrown, such
as a crash of MATLAB, is caught in the \mcode{catch} statement. On
the continuous integration server, this leads to a returned exit code
of \mcode{1}, while on a user computer, the exception is explicitly
rethrown.

The core of the continuous integration system is the test suite. In
other words, the number and quality of tests determine the quality
of the code in the \emph{/src }directory. Writing tests may be a long
process, in particular for bug prone code. In the case of \texttt{the
COBRA Toolbox}, the community picked up writing tests for their own
functions, leading to a steady increase of code quality (see Section
\ref{subsec:Code-quality}).

% --------------------------------------------------------------------------------
\subsection{Continuous integration feedback}
\label{subsec:Continuous-integration-feedback}
\label{subsec:Build-status}

An execution of a job on the continuous integration server, or a run
of the MATLAB test suite, is considered as a \emph{build}. 
For each commit on the \emph{develop} and \emph{master}
branches, a job- and build-specific build status is set on GitHub and updated
continuously while the build is running. This commit build status reflects the actual
status of the build on the server. The build status is 
either a \emph{green check mark} for \emph{success}, a \emph{red cross
mark} for \emph{failure}, or a \emph{yellow dot} for a job that is
pending. A commit can hence cause
the failure or the success of a job. The trigger build status, or global build
status, is set as successful when all jobs ran without errors on a
certain operating system.

Commonly, one or multiple visible badges are displayed visibly on
the first page of a GitHub repository that indicate the status of
the latest builds of the \emph{develop} branch. The badges are a visual
aid of quickly determining the stability and reproducibility of the
development of a repository.

For the case of \texttt{the COBRA Toolbox} repository, the MATLAB
test suite is run on multiple operating systems. As for each build
(i.e., for each MATLAB version), a build status is returned, a matrix
of build status is defined, which can be consulted under \href{http://opencobra.github.io/cobratoolbox/docs/builds.html}{opencobra.github.io/cobratoolbox/docs/builds.html}.
In order to simplify the readout for the end-user who is primarily
interested in the stability of \texttt{the COBRA Toolbox} on a specific
platform, a build status is retrieved for each operating system. To
this end, a Julia script that retrieves continuously the list of build
status for the last commit on the \emph{develop} branch using \texttt{Github.jl}
(\href{http://github.com/JuliaWeb/GitHub.jl}{github.com/JuliaWeb/GitHub.jl}),
which provides a Julia interface to the GitHub API v3. The overall
status of the build is determined for each operating system, and a
visual badge is set on the \texttt{Jenkins} web server. In the rare
case of a build failing because of a misconfiguration of a continuous
integration server node or in case of emergency, a build status of
a commit may also be changed manually using the GitHub API v3.

\begin{figure*}[t]
\noindent \begin{centering}
\includegraphics[width=1\textwidth]{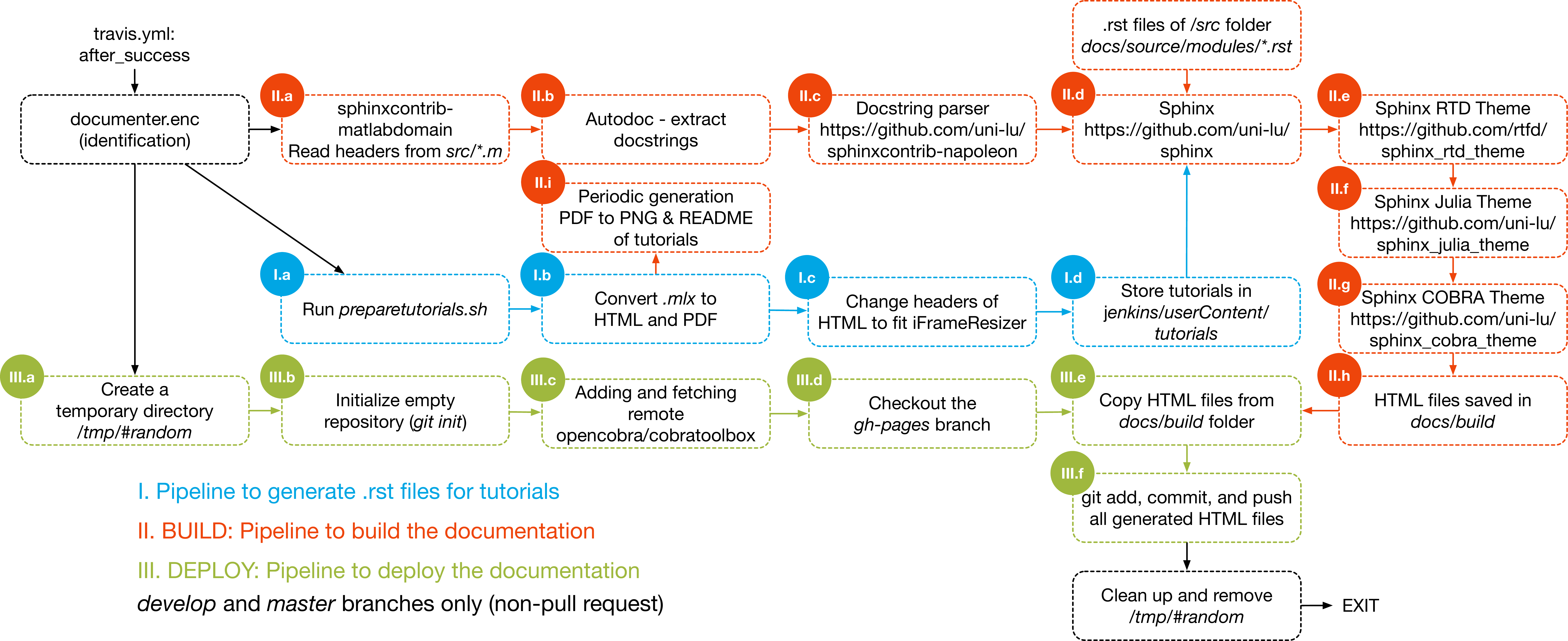}
\par\end{centering}
\vspace{-1mm}
\caption{Pipeline for generating, building,
and deploying the documentation using \texttt{Documenter.py}. The
documentation is automatically deployed from the continuous integration
server and generated based on the function headers.}
\label{fig:pipelineDocumentation}
\end{figure*}

Another valuable feedback mechanism is to alert the administrator
of the continuous integration system of build failures by email. Other
health monitoring tools, such as explained in Section \ref{subsec:Virtual-machine-management},
help determine the cause of failure in cases where the failure of
a build might be system related (e.g., high memory usage or faulty
disks).

\vspace{-5mm}
% --------------------------------------------------------------------------------
\section{Documentation and tutorials}
\label{sec:Documentation-and-tutorials}

The documentation of \texttt{the COBRA Toolbox} is generated and deployed
automatically once a push is made on the\emph{ master} or \emph{develop}
branch of the GitHub repository, and if the build is successful. 

The building of the documentation is done in 3 phases: the generation
of the tutorials, the generation of the documentation and its deployment
to a host server. Figure \ref{fig:pipelineDocumentation} represents
all the steps involved in the creation and deployment of tutorials
and documentation, and summarises the creative pipelines. 

As MATLAB does not provide a full documentation system to publish
documentation of user created MATLAB codes, a specific documentation
pipeline that uses the popular Python documentation tool \emph{Sphinx}
(\href{http://www.sphinx-doc.org/en/stable/index.html}{sphinx-doc.org})
has been developed. \texttt{Documenter.py} (\href{https://github.com/syarra/Documenter.py}{github.com/syarra/Documenter.py})
is at the heart of the documentation pipeline. The package is designed
to combine\emph{ reStructuredText} files and inline \textit{docstrings}
from MATLAB functions into a single interlinked and publishable documentation.
A \textit{docstring} of a function is a string literal specified in
source code that is used to document a function in an easily comprehensible
way. In order for the documentation to be generated properly, the
docstrings must be written using keywords listed in Table~\ref{tab:keywords-documentation}.A keyword defines the start of a block of documentation with the header
of a function, must be followed by non-empty lines, and is separated
from the next block by an empty line. 
\vspace{-5mm}
\begin{table}[H]
\begin{centering}
\small{%
\begin{tabular*}{1\columnwidth}{@{\extracolsep{\fill}}l|>{\raggedright}p{5cm}}
\textbf{Keyword} & \textbf{Purpose}\tabularnewline
\hline 
\mcode{USAGE:} & defines how to use the function\tabularnewline
\hline 
\mcode{INPUT:} or \mcode{INPUTS:} & describes input argument(s)\tabularnewline
\hline 
\mcode{OUTPUT:} or \mcode{OUTPUTS:} & describes the output argument(s) \tabularnewline
\hline 
\mcode{EXAMPLE:} & shows example of code (MATLAB syntax)\tabularnewline
\hline 
\mcode{NOTE:} & displays a highlighted box with text\tabularnewline
\hline 
\mcode{AUTHOR:} & lists author(s) of the function\tabularnewline
\hline 
\end{tabular*}}
\par\end{centering}
\vspace{1mm}
\caption{\label{tab:keywords-documentation}Main keywords in the function headers
(docstrings) used for documentation extraction.}
\vspace{-3mm}
\end{table}

The first phase in the generation of the full documentation
is the generation of tutorials (steps \textbf{I.a}-\textbf{I.d} in
Figure \ref{fig:pipelineDocumentation}). MATLAB provides a very convenient
way of writing tutorials using\emph{ Live Script}, which is a document
containing both computer code and text elements, such as paragraphs,
equations, figures, or links, and which is saved as a \emph{.mlx}
file. In order to allow for swift consultancy of the tutorial through
a web browser, or even print a hard copy of the tutorial, the \emph{Live
Scripts} are converted automatically into PDF and HTML formats by
executing the \emph{prepareTutorials.sh} script (steps\textbf{ I.a
}and\textbf{ I.b}) based on MATLAB functionality. During steps\textbf{
I.c} and \textbf{I.d}, the \emph{.pdf} and \emph{.html} versions of
the tutorials are further modified to fit the webpage style and moved
to the web server location. A user running MATLAB R2016a or above
may consult the tutorials in 3 different formats: as a document (\emph{.pdf}),
on the web (\emph{.html}), or locally directly in MATLAB (\emph{.mlx}).
The Live Script functionality is not available for versions of MATLAB
older than R2016a. For convenience, the\emph{ .pdf }documents are
converted to \emph{.png} image files and displayed directly within
the README.md files on GitHub when browsing the \emph{/tutorials }directory
(\href{https://github.com/opencobra/cobratoolbox/tree/master/tutorials}{github.com/opencobra/cobratoolbox/tree/master/tutorials}).

The second phase consists of extracting the docstrings from MATLAB
functions using the \texttt{Sphinx} package and the \texttt{matlabdomain}
plugin (\href{https://pypi.python.org/pypi/sphinxcontrib-matlabdomain}{pypi.python.org/pypi/sphinxcontrib-matlabdomain})
(steps \textbf{II.a}-\textbf{II.d}) based on the docstrings and the
keywords shown in Table \ref{tab:keywords-documentation}. The docstrings
are then combined with the \emph{reStructuredText} files located in
the \textit{/docs/source/modules} directory from the original repository
to produce HTML pages (steps \textbf{II.e}-\textbf{II.h}) that can
be displayed on the web server. Style and layout are matched to the
style of the web documentation for Julia packages thanks to the \texttt{Sphinx}
COBRA Theme package (\href{https://github.com/uni-lu/sphinx_cobra_theme}{github.com/uni-lu/sphinx\_{}cobra\_{}theme}).
The Julia Sphinx theme (\href{https://github.com/uni-lu/sphinx_julia_theme}{github.com/uni-lu/sphinx\_{}julia\_{}theme})
is adopted in order to provide a harmonised package suite together
with \texttt{COBRA.jl} \citep{Hei17} (\href{https://opencobra.github.io/COBRA.jl/stable}{opencobra.github.io/COBRA.jl/stable}).

The third and last phase of the documentation generation is the deployment
to the host web server (\href{https://opencobra.github.io/cobratoolbox}{opencobra.github.io/cobratoolbox})
via the \emph{gh-pages }branch of \texttt{the COBRA Toolbox} repository
(\href{https://github.com/opencobra/cobratoolbox/tree/gh-pages}{github.com/opencobra/cobratoolbox/tree/gh-pages}).
As explained in Section \ref{subsec:Local-and-Github},
the user \emph{cobrabot} pushes the newest changes and publishes the
latest version of the documentation.

% --------------------------------------------------------------------------------
\section{Evaluation of code quality}
\label{subsec:Code-quality}

% --------------------------------------------------------------------------------
\subsection{Code coverage}
\label{subsec:Code-coverage}

Functional coverage provides information about which scenarios have
been tested. In software development, it is essential to track test
coverage, or in other words, which functions and source lines are
executed during the test run. The coverage report helps to estimate
how much of the code base is tested or executed without crashing and
how much code is not covered through testing. Code coverage is reported
through a coverage report generator for MATLAB and GNU Octave, namely
the \texttt{MOcov} toolbox (\href{http://github.com/MOcov/MOcov}{github.com/MOcov/MOcov}),
and through the free and GitHub integrated \href{https://codecov.io/}{codecov.io}
service. A code coverage report reveals which lines have been added
and whether they have been tested. The code coverage difference can
consequently be determined for every pull request.

The code coverage determines the scope of the test suite and is determined
as a ratio of the number of executed code source lines and the total
number of executable lines of code. Executable lines of code are counted
as all lines in the\emph{ .m} files in the \emph{/src} folder that
do not start with \mcode{\%} or the language keywords \mcode{end},
\mcode{otherwise}, \mcode{switch}, \mcode{else}, \mcode{case},
or \mcode{function}.

Tracking the code coverage provides a measure of stability of the
code base and the breadth of the test suite. The limitations of code
coverage reports include that the intended functionality of the code
is not verified and that the quality of the code is not assessed.
Although there is no precise measure for code quality itself, the
efficiency of code certainly can be graded (see Section \ref{subsec:Code-quality-grade}).

% --------------------------------------------------------------------------------
\subsection{Code efficiency grade}
\label{subsec:Code-quality-grade}

The code efficiency grade is a valuable measure of how MATLAB code
can be improved for efficiency and to detect potential problems and
opportunities for code improvement. The implemented function \mcode{checkcode}
is run on each source code file and the number of MATLAB \texttt{Code
Analyzer} messages is recorded. 
%\vspace{-3mm}
\begin{table}[H]
\noindent \begin{centering}
\small{%
\begin{tabular}{cc}
\textbf{Code efficiency grade} & \textbf{Percentage range}\tabularnewline
\hline 
A & 0\% - 3\%\tabularnewline
\hline 
B & 3\% - 6\%\tabularnewline
\hline 
C & 6\% - 9\%\tabularnewline
\hline 
D & 9\% - 12\%\tabularnewline
\hline 
E & 12\% - 15\%\tabularnewline
\hline 
F & > 15\%\tabularnewline
\hline 
\end{tabular}}
\par\end{centering}
\vspace{1mm}
\caption{Conversion chart from coverage percentage to quality grade.}
\label{tab:Code-grade-percentage}
\end{table}
\vspace{-3mm}
The average number of messages per source code file is divided by
the actual number of executable source code lines, which yields the
code grade percentage. For ease of use, the code grade percentage
is converted to a letter code as shown in Table \ref{tab:Code-grade-percentage}.

% --------------------------------------------------------------------------------
\subsection{Code linting}

Linting is the practice of harmonising a given code style across an
entire code base. The need for code linting of an open-source code
base is obvious; many developers contributing to the same project
and source files may not have the same discipline of writing code
according to style guidelines. 

A clean, homogeneous, and easily readable code base is required for
easy debugging and swift maintenance. \texttt{Automatlab.py} (\href{https://github.com/syarra/automatlab}{github.com/syarra/automatlab})
is a tool designed to ensure that MATLAB code comply to style conventions
defined at \href{https://opencobra.github.io/cobratoolbox/docs/styleGuide.html}{opencobra.github.io/cobratoolbox/docs/styleGuide.html}.
\texttt{Automatlab.py} runs in two steps: first, a compliance analysis
is performed to check the adherence to the style guide. Then, formatting
is fixed directly in the MATLAB code.

Linting is a common practice when coding using Python (e.g., the package
\texttt{AutoPEP8}: \href{http://github.com/hhatto/autopep8}{github.com/hhatto/autopep8})
or Julia (e.g., the package \texttt{Lint.jl}: \href{http://github.com/tonyhffong/Lint.jl}{github.com/tonyhffong/Lint.jl}).
Several state-of-the art editors, such as \emph{Atom} (\href{https://atom.io/}{atom.io}),
use built-in linting functionality to advise the developer of best
coding practices. \hfill{}$\blacksquare$

% --------------------------------------------------------------------------------
\section*{List of Acronyms}
\vspace{-3mm}
\noindent \begin{center}
\begin{tabular*}{1\columnwidth}{@{\extracolsep{\fill}}>{\raggedright}p{2cm}l}
\textbf{Acronym} & \textbf{Designation}\tabularnewline
\hline 
ARTENOLIS & Automated Reproducibility and Testing Environment\\
&  for Licensed Software\tabularnewline
\hline 
COBRA & COnstraint-Based Reconstruction and Analysis\tabularnewline
\hline 
HTTP & Hypertext Transfer Protocol\tabularnewline
\hline 
JNLP & Java Network Launch Protocol \tabularnewline
\hline 
KVM & Kernel-based Virtual Machine\tabularnewline
\hline 
NFS & Network File System\tabularnewline
\hline 
PR & Pull request\tabularnewline
\hline 
SHA1 & Secure Hash Algorithm 1\tabularnewline
\hline 
SMB & Server Message Block\tabularnewline
\hline 
SSH & Secure Shell \tabularnewline
\hline 
YAML & YAML Ain't Markup Language\tabularnewline
\hline 
\end{tabular*}
\par\end{center}

% --------------------------------------------------------------------------------
\bibliographystyle{bioinformatics}
\fontsize{7}{9}\selectfont

\end{multicols}

\end{document}